\documentclass[aps, reprint, nofootinbib, showpacs, a4paper]{revtex4-1}

\usepackage{amssymb}
\usepackage[intlimits]{amsmath}

\begin{document}

\title{Macroscopic detection of deformed QM by the harmonic oscillator}

\author{Michael~Maziashvili }
\email{maziashvili@iliauni.edu.ge} \affiliation{School of Natural Sciences and Engineering, Ilia State University,\\ 3/5 Cholokashvili Ave., Tbilisi 0162, Georgia}

\begin{abstract}
Based on the nonperturbative analysis, we show that the classical motion of harmonic oscillator derived from the deformed QM is manifestly in contradiction with observations. For this reason, we take an alternate way for estimating the effect and discuss its possible observational manifestations in macrophysics.         
\end{abstract}


%04.60.Bc Phenomenology of quantum gravity

\maketitle

Motivated by the paper \cite{Pikovski:2011zk} (see also \cite{Pikovski:2014uua}), suggesting a way for observing the Planck length deformed QM via the optomechanical experiments, we take an attempt to discuss further this question. This paper provides the possible experimental  basis for observing the corrections to the QM, say, of the form (we employ units in which: $c=\hbar =1$) 

\begin{equation}\label{Planck-Skala QM}
\left[\widehat{X}, \widehat{P}_X \right] \,=\, i\left(1 \,+\, \ss \widehat{P}_X^2\right)  ~,
\end{equation} by the interference effect when the part of the light entering the optomechanical device is subject to the oscillation of the moving mirror-wall of the device, under assumption that the latter undergoes "quantum motion" with respect to the modified prescription \eqref{Planck-Skala QM}. In Eq.\eqref{Planck-Skala QM} $\ss \simeq l_P^2$, where $l_P$($\simeq 10^{-33}$cm) is the Planck length. It it naturally expected that the gravitational corrections to the processes probing the length scale $\ell$($ \gg l_P$) should be suppressed by some powers of the ratio $\l_P/\ell$ \cite{Donoghue:2015hwa}. The Eq.\eqref{Planck-Skala QM} can be interpreted in much the same way. Namely, one may hold that because of quantum fluctuations of the background metric - any length scale, $\ell \gg l_P$, acquires uncertainty that on the general grounds can be parameterized as: $\simeq l_P^\alpha\ell^{1-\alpha}$. Hence the position uncertainty of the particle gets increased, $\delta x \rightarrow \delta x + \ss^{\alpha/2}\delta x^{1-\alpha}$, leading to the modified uncertainty relation $\delta x\delta p \gtrsim 1+ \left(\sqrt{\ss}\delta p\right)^\alpha$. The Eq.\eqref{Planck-Skala QM} corresponds to the specific case: $\alpha =2$. Another case of interest might be: $\alpha=1/2$ \cite{Maziashvili:2012dd}. We take this judgment as a guiding principle for our further discussion. It can be stated simply as follows: the size of the gravitational corrections depends primarily on the length scale probed by the system.

Returning to the discussion of \cite{Pikovski:2011zk, Pikovski:2014uua}, the system under consideration consists of a monochromatic electromagnetic field confined in a mirror box - one of the walls of which is vibrating non-relativistically. The vibrating wall, which is assumed to be parallel to the $(Y, Z)$ plane, is moving along the $X$ axis. This system can be modeled by the Hamiltonian

\begin{equation}\label{Hamiltonoperator}
\widehat{H} \,=\,  \omega_{ph} \widehat{a}\,^+\widehat{a} \,+\, \frac{\widehat{\mathbf{P}}\,^2}{2m} \,+\, \frac{m\omega^2 \widehat{X}^2}{2} \,-\,  g\widehat{a}\,^+\widehat{a} \widehat{X} ~,
\end{equation} where the three-dimensional momentum $\mathbf{P}$ (instead of the one-dimensional one) is introduced purely for technical convenience,  $\omega_{ph}$ is the photon frequency and $\omega$ and $\widehat{X}$ stand for the vibrating mirror-wall frequency and position operator, respectively\footnote{Detailed derivation of this Hamiltonian in one-dimensional case can be found in \cite{Moore, Law:1995zz}.}. The coupling parameter $g$ can be elucidated in terms of input parameters in the following way. Because of photon gas pressure, which is equal to the one third of the photon gas energy density, under assumption that the amplitude of mirror vibrations is much smaller then the cavity length, that is, $X \ll l$, there is in fact a constant force acting in outward direction on the moving wall 

\begin{equation}
\mathcal{F}_X \,=\, \frac{\omega_{ph}n_{ph}s}{3ls} ~,  \nonumber
\end{equation} where $s$ denotes its surface area and $l$ stands for the cavity length. Hence the coupling parameter: $g = \omega_{ph}/3l$. 

One can construct a particular representation for the deformed $\widehat{X}, \widehat{Y}, \widehat{Z}, \, \widehat{\mathbf{P}}$ operators in terms of the standard position and momentum operators, $\widehat{x}, \widehat{y}, \widehat{z}, \, \widehat{\mathbf{p}}$, for an arbitrary $\alpha$ ($\alpha \neq 1$) \cite{Maziashvili:2012dd}: $\widehat{X} \,=\, \widehat{x}, \widehat{Y} \,=\, \widehat{y}, \widehat{Z} \,=\, \widehat{z}$,  

\begin{equation}
\widehat{\mathbf{P}} \,=\, \widehat{\mathbf{p}}\left[1 \,-\, (\alpha-1)\left(\ss \widehat{\mathbf{p}}^2\right)^{\alpha/2}  \right]^{\frac{1}{1-\alpha}} ~. \nonumber
\end{equation} When $\alpha=2$, one arrives at a well known result of \cite{Kempf:1996nk}. It is worth noticing that the case $\alpha > 1$ ($\alpha < 1$) corresponds to the assumption that the length fluctuation $\delta \ell = \ss^{\alpha/2}\ell^{1-\alpha}$ decreases (increases) with increasing of $\ell$. A specific feature for the deformed QM with $\alpha > 1$ is that in this case the state vectors are represented by the cutoff Fourier representation\footnote{The interested reader may be referred to \cite{Maziashvili:2012zr}.}, but it is less important for us as we are dealing with the objects having the localization with much greater than the Planck length.

By using this particular representation, the Hamiltonian \eqref{Hamiltonoperator} gets modified as

 \begin{eqnarray}\label{Hamiltonian} \widehat{H} \,=\,  \omega_{ph} \widehat{a}\,^+\widehat{a} \,+\, \frac{\widehat{\mathbf{p}}\,^2\left[1 \,-\, (\alpha-1)\left(\ss \widehat{\mathbf{p}}^2\right)^{\alpha/2}  \right]^{\frac{2}{1-\alpha}}}{2m} \,+\, \nonumber \\ \frac{m\omega^2 \widehat{x}^2}{2} \,-\,  g\widehat{a}\,^+\widehat{a} \widehat{x} ~.~~
\end{eqnarray} The Heisenberg equations of motion read

\begin{widetext}
\begin{eqnarray}\label{modzraobisgantolebebi}
\dot{\widehat{a}} \,=\, -i\omega_{ph} \, \widehat{a} \,+\, ig\, \widehat{a} \, \widehat{x} ~,  ~~ \dot{\widehat{p}_x} \,=\, -m\omega^2 \widehat{x} \,+\, g \widehat{a}^+\widehat{a} ~, ~~ \dot{\widehat{p}_y} \,=\, \dot{\widehat{p}_z} \,=\, 0 ~, \\\label{gantolebaxistvis}  \dot{\widehat{x}} \,=\, \frac{\widehat{p}_x\left[1 \,-\, (\alpha-1)\left(\ss \widehat{\mathbf{p}}^2\right)^{\alpha/2}  \right]^{\frac{2}{1-\alpha}}}{m} \,+\, \frac{\alpha \widehat{p}_x\left(\ss \widehat{\mathbf{p}}^2\right)^{\alpha/2}\left[1 \,-\, (\alpha-1)\left(\ss \widehat{\mathbf{p}}^2\right)^{\alpha/2}  \right]^{\frac{1+\alpha}{1-\alpha}}}{m}   ~. 
\end{eqnarray}\end{widetext} The equations for $\widehat{y}(t)$ and $\widehat{z}(t)$ have plainly the same form as Eq.\eqref{gantolebaxistvis} with $\widehat{p}_x$ replaced by $\widehat{p}_y$ and $\widehat{p}_z$, respectively. The solution for $\widehat{a}(t)$ can immediately be written as  

\begin{eqnarray}\label{elmagveli}
\widehat{a}(t) \,=\, \widehat{a}(0) \, \exp\left(-i\omega_{ph} t \,+\, ig\int_0^t\mathrm{d}\tau \widehat{x}(\tau)\right) ~.
\end{eqnarray} From this solution it is clear that $\widehat{a}^+(t)\widehat{a}(t)=\widehat{a}^+(0)\widehat{a}(0)$, which enables one to rewrite the equation for $\widehat{p}_x(t)$ in the form 

\begin{equation}\label{mesamegantoleba}
 \dot{\widehat{p}}_x \,=\, -m\omega^2 \widehat{x} \,+\, g \widehat{a}^+(0)\widehat{a}(0) ~.
\end{equation} 

Let us now specify the initial state of the system. For this purpose, we take the coherent state for the photon gas $\widehat{a}(0)|\zeta\rangle = \zeta|\zeta\rangle$, and the minimum uncertainty state\footnote{Here we mean the uncertainty relation between the standard position and momentum operators.} \begin{eqnarray}\label{minganuzghvreli}
\psi(x,\,y,\,z) \,=\, \exp\left(ip_x(0)x \,-\, \frac{\left[x-x(0)\right]^2}{4(\delta x)^2} \right)\times \nonumber \\ \frac{1}{\sqrt[4]{(2\pi)^3}\sqrt{\delta x\delta y\delta z}} \, \exp\left(-\, \frac{y^2}{4(\delta y)^2} -\, \frac{z^2}{4(\delta z)^2} \right) ~. \nonumber ~~~~~~
\end{eqnarray} for the moving mirror. For averaged quantities with respect to these states one gets the following equations  

\begin{eqnarray}
\dot{p}_x \,=\, -m\omega^2 x \,+\, g |\zeta|^2~,~ p_y \,=\, p_z \,=\, 0 ~, ~~ y \,=\, z \,=\, 0 ~, \nonumber \\   \dot{x} \,=\, \frac{p_x\left[1 \,-\, (\alpha -1) \left(\ss p_x^2\right)^{\alpha/2}  \right]^{\frac{2}{1-\alpha}}}{m} \,+\, ~~~~~~~~~~~~~~~ \nonumber \\ \frac{\alpha p_x\left(\ss p_x^2\right)^{\alpha/2}\left[1 \,-\, (\alpha-1) \left(\ss p_x^2\right)^{\alpha/2}  \right]^{\frac{1+\alpha}{1-\alpha}}}{m} ~.~~~ \label{sichkare}
\end{eqnarray} These equations can be derived from classical Hamiltonian 

\begin{eqnarray}\label{ertganzomilebiani}
H \,=\, \frac{p^2_x\left[1 \,-\, (\alpha-1)\left(\ss p_x^2\right)^{\alpha/2}  \right]^{\frac{2}{1-\alpha}}}{2m} \,+\, \nonumber \\ \frac{m\omega^2 x^2}{2} \,-\, g|\zeta|^2x ~. ~~
\end{eqnarray} Let us notice that by adding a linear term to the potential of oscillator - the potential has still the pure oscillator form with the equilibrium position shifted. So, this term in Eq.\eqref{ertganzomilebiani} can merely be dropped as in the paper \cite{Bawaj:2014cda}. The question of experimental detection can be addressed merely by measuring the velocity $\dot{x}$ for $x=0$. The conservation of energy allows one to estimate the value of momentum $p_x$ for $x=0$,   

\begin{equation}\label{impulsi}
\frac{p^2_x\left[1 \,-\, (\alpha-1)\left(\ss p_x^2\right)^{\alpha/2}  \right]^{\frac{2}{1-\alpha}}}{2m}  \,=\, \frac{m\omega^2x^2_{max}}{2} ~, 
\end{equation} and then by substituting this value of $p_x$ in \eqref{sichkare} - to determine the velocity at this point. It is almost obvious that the deviation from the standard velocity becomes appreciable when the momentum approaches $\ss^{-1/2}$ (Planck energy scale). Let us consider two specific cases $\alpha = 2$ and $\alpha=1/2$. In the former case, assuming $|p_x|$ is approaching $\ss^{-1/2}$ ($|p_x| < \ss^{-1/2}$) and usig Eq.\eqref{impulsi}, one obtains for the velocity at $x=0$ 

\begin{eqnarray}
m|\dot{x}| \,=\, |p_x| \, \frac{1+\ss p_x^2}{\left(1-\ss p_x^2\right)^3}= \left(m\omega x_{max} \right)^3\left(\ss + \frac{1}{p_x^2} \right) \nonumber \\  \simeq 2\ss \left(m\omega x_{max} \right)^3 ~. \nonumber 
\end{eqnarray} Now let us turn to the case $\alpha =1/2$. With no loss of generality, we can absorb the factor $1-\alpha$ in $\ss^{\alpha/2}$ (for $\ss$ is defined up to the numerical factor of order unity). Then under assumption $\ss p_x^2 \simeq 1$ at $x=0$, from Eq.\eqref{impulsi} for for the velocity (at $x=0$) one obtains  

\begin{eqnarray}
m\omega x_{max} \simeq 4/\sqrt{\ss} ~, ~~
m|\dot{x}| = |p_x| \left(1+ \left(\ss p_x^2\right)^{1/4} \right)^3 \times \nonumber \\ \left(1+ 2\left(\ss p_x^2\right)^{1/4} \right) \simeq \frac{24}{ \sqrt{\ss}} ~. \nonumber
\end{eqnarray} For characterizing the size of the effect - it might be useful to look at the ratio $|\dot{x}|/\omega x_{max}$, which in the standard case is equal to $1$. Such a big deviation at the macroscopic scales is not natural\footnote{This point can easily be missed in the framework of perturbative treatment \cite{Bawaj:2014cda} the validity of which requires $\left(\ss\mathbf{p}^2\right)^{\alpha/2}\ll 1$.} that leads to the opinion to revise the whole discussion from the very outset.

For this purpose, let us look at the well known QM description of the classical motion of harmonic oscillator \cite{Landau}. The wave function

\begin{eqnarray}\label{talghurifunktsia}
\psi(t, x) \,=\, \sqrt[4]{\frac{m\omega}{\pi}}\exp\left\{ip_x(t)x \,-\, \frac{m\omega(x-x(t))^2}{2} \,-\, \right. \nonumber \\ \left.  \frac{i\omega t}{2} \,-\, \frac{ip_x(t)x(t)}{2} \right\} ~, ~~
\end{eqnarray} where $x(t)$ and $p_x(t)$ obey the classical equations for harmonic oscillator,

\begin{equation}
\dot{p}_x(t) \,=\, -m\omega^2 x(t) ~, ~~ \dot{x}(t) \,=\, \frac{p_x(t)}{m} ~, \nonumber
\end{equation} solves the Schr\"odinger equation for harmonic oscillator. It is plain to see that: $p_x(t) = \langle\psi(t)|\widehat{p}_x|\psi(t)\rangle, \, x(t) = \langle\psi(t)|\widehat{x}|\psi(t)\rangle$. On the other hand, this wave function corresponds to the minimum uncertainty state: $\delta x= 1/\sqrt{2m\omega}, \, \delta p_x = \sqrt{m\omega/2}$. The classicality of motion means that the quantum fluctuations should be smaller as compared to the classical quantities. It is convenient to characterize the degree of classicality in terms of the energy 

\begin{eqnarray}
\langle\psi(t)|\text{Oscillator Hamiltonian}|\psi(t)\rangle \,=\,   \frac{p_x^2(t)}{2m} \,+\,  \frac{m\omega^2x^2(t)}{2}  \nonumber \\  \,+\, \frac{(\delta p_x)^2}{2m} \,+\, \frac{m\omega^2(\delta x)^2}{2} \,=\,   \frac{p_x^2(t)}{2m} \,+\, \frac{m\omega^2x^2(t)}{2} \,+\, \frac{\omega}{2} ~. \nonumber 
\end{eqnarray} Expanding $|\psi(t)\rangle$ into the {\tt Eigenfunktionen}, $|j\rangle$, of harmonic oscillator Hamiltonian

\begin{equation}
|\psi(t)\rangle \,=\, \exp\left(-\frac{i\omega t}{2} -\frac{|b(t)|^2}{2}\right)\sum\limits_{j=0}^\infty \frac{b^j(t)}{\sqrt{j!}} \, |j\rangle ~, \end{equation} where \begin{equation} b(t) \,=\, \frac{m\omega x(t) \,+\, ip_x(t)}{\sqrt{2 m\omega}} ~, \nonumber
\end{equation} the energy can be expressed in terms of the mean occupation number, $n$, as 

\begin{equation}
\frac{p_x^2(t)}{2m} \,+\, \frac{m\omega^2x^2(t)}{2} \,=\, \omega n ~. \nonumber
\end{equation} Therefore, the motion is essentially classical if $n \gg 1$. This discussion tells us that there is an oscillating particle localized within the region $\delta x \sim 1/\sqrt{m\omega}$ and for estimating the gravitational corrections to this system with respect to physics underlying the Eq.\eqref{Planck-Skala QM}, one has to use this length scale and look for the corrections in the form $(l_P/\delta x)^2$. So that, the size of the gravitational correction depends primarily on $\delta x$. Certainly, the moving mirror in optomechanical device with the parameters given in \cite{Pikovski:2011zk}, $m\simeq M_P, \, \omega \simeq 10^{5}$Hz, is not localized within the region $1/\sqrt{M_P \omega}\sim 10^{-14}$cm but for the moment let us forget about it and ask what might be the size of the gravitational corrections to the classical oscillator in the framework of the above discussion. As it was discussed in the beginning of the paper, because of fluctuating background the oscillator gets modified in such a way as to respect the modified uncertainty relation
\begin{equation}
\delta x\delta p_x \,\geq \, \frac{1}{2} \left\{1 \,+\, \left(\sqrt{\ss}/\delta x\right)^{\alpha}\right\} \,=\, \frac{1}{2} \left\{1 \,+\, \left(\ss m\omega\right)^{\alpha/2}\right\} ~. \nonumber
\end{equation} Let us continue in the spirit of the previous discussion and ascribe this modification to the momentum operator: $\widehat{p}_x\rightarrow \widehat{p}_x \left[1 \,+\, \left(\sqrt{\ss}/\delta x\right)^\alpha \right]$. It makes sense because for the problem under consideration $\delta x$ is constant. Thus, we are just saying that the momentum for oscillating particle, which is localized within the region $\delta x$, is increased by the factor: $1+\left(\sqrt{\ss}/\delta x\right)^{\alpha}$. The modified Hamiltonian 

\begin{eqnarray}
 \frac{\left[1 + \left(\sqrt{\ss}/\delta x\right)^\alpha\right]^2\widehat{p}_x^2}{2m} + \frac{m\omega^2\widehat{x}^2}{2} =  \frac{\widehat{p}^2_x}{2m'} + \frac{m'\omega'^2\widehat{x}^2}{2} ~, \nonumber  \\ m' = \frac{m}{\left[1 + \left(\sqrt{\ss}/\delta x\right)^\alpha\right]^2} ~, ~~~ \omega' = \omega \left[1 + \left(\sqrt{\ss}/\delta x\right)^{\alpha}\right] ~, \nonumber
\end{eqnarray} in view of the above discussion, tells us that the classical motion gets modified as      

\begin{equation}
x(t) \,=\, x(0)\cos(\omega't) ~, ~~ p_x(t) \,=\, -m'\omega'x(0)\sin(\omega't) ~. \nonumber
\end{equation} In the case when the localization width of the oscillating body, $\ell$, is much bigger than $\delta x= 1/\sqrt{2m\omega}$, one could make a straightforward generalization of this result 

\begin{eqnarray}
 x(t) &=& x(0)\cos\left( \left[1 \,+\, \left(\sqrt{\ss} /\ell\right)^\alpha\right]\omega t\right) ~.   \nonumber 
\end{eqnarray} Obviously, it is extremely hard to detect such a deviation from standard picture for the realistic input parameters. Even if one takes an extreme value: $\ell \sim 10^{-8}$cm, the correction term $\left(\sqrt{\ss} /\ell\right)^\alpha$ becomes $\sim 10^{-12}$ for $\alpha=1/2$ and $\sim 10^{-50}$ for $\alpha = 2$.

Summarizing, we see that a classical motion derived from the nonperturbative treatment of quantum oscillator on the basis of deformed QM \eqref{Planck-Skala QM} is almost obviously in conflict with "macrophysics". This point has already been stressed in  \cite{Maziashvili:2012zr}, in the framework of somewhat different nonperturbatie analysis, and also indicated in \cite{Silagadze:2009vu} - in the framweork of perturbative treatment. Nevertheless, an alternative treatment guided by the physics that underlies the \eqref{Planck-Skala QM} leads to the result that hardly can be detected via the macrophysics.

\acknowledgments
Useful comments from Zurab Silagadze and David Vitali are kindly acknowledged. The research was partially supported by the Shota Rustaveli National
Science Foundation under contract number 31/89.


\begin{thebibliography}{10}




%\cite{Pikovski:2011zk}
\bibitem{Pikovski:2011zk} 
  I.~Pikovski, M.~R.~Vanner, M.~Aspelmeyer, M.~S.~Kim and C.~Brukner,
  %``Probing Planck-scale physics with quantum optics,''
  Nature Phys.\  {\bf 8}, 393 (2012)
  [arXiv:1111.1979 [quant-ph]].
  %%CITATION = ARXIV:1111.1979;%%
  %72 citations counted in INSPIRE as of 14 Oct 2015



%\cite{Pikovski:2014uua}
\bibitem{Pikovski:2014uua} 
  I.~Pikovski, PhD Thesis "Macroscopic quantum systems and gravitational phenomena", (URL: http://inspirehep.net/record/1365384).
  %%CITATION = INSPIRE-1365384;%%



%\cite{Donoghue:2015hwa}
\bibitem{Donoghue:2015hwa} 
  J.~F.~Donoghue and B.~R.~Holstein,
  %``Low Energy Theorems of Quantum Gravity from Effective Field Theory,''
  J.\ Phys.\ G {\bf 42}, no. 10, 103102 (2015)
  %doi:10.1088/0954-3899/42/10/103102
  [arXiv:1506.00946 [gr-qc]].
  %%CITATION = doi:10.1088/0954-3899/42/10/103102;%%
  %4 citations counted in INSPIRE as of 24 Dec 2015


%\cite{Maziashvili:2012dd}
\bibitem{Maziashvili:2012dd} 
  M.~Maziashvili,
  %``Field propagation in a stochastic background space: The rate of light incoherence in stellar interferometry,''
  Phys.\ Rev.\ D {\bf 86}, 104066 (2012)
  %doi:10.1103/PhysRevD.86.104066
  [arXiv:1206.4388 [gr-qc]].
  %%CITATION = doi:10.1103/PhysRevD.86.104066;%%
  %5 citations counted in INSPIRE as of 24 Dec 2015




\bibitem{Moore}
G.~T.~Moore, 
% "Quantum Theory of the Electromagnetic Field in a Variable‐Length One‐Dimensional Cavity",
  J.\ Math.\ Phys. {\bf 11}, 2679 (1970).

  

%\cite{Law:1995zz}
\bibitem{Law:1995zz} 
  C.~K.~Law,
  %``Interaction between a moving mirror and radiation pressure: A Hamiltonian formulation,''
  Phys.\ Rev.\ A {\bf 51}, 2537 (1995).
  %%CITATION = PHRVA,A51,2537;%%
  %51 citations counted in INSPIRE as of 15 juin 2015





%\cite{Kempf:1996nk}
\bibitem{Kempf:1996nk} 
  A.~Kempf and G.~Mangano,
  %``Minimal length uncertainty relation and ultraviolet regularization,''
  Phys.\ Rev.\ D {\bf 55}, 7909 (1997)
  %doi:10.1103/PhysRevD.55.7909
  [hep-th/9612084].
  %%CITATION = doi:10.1103/PhysRevD.55.7909;%%
  %251 citations counted in INSPIRE as of 24 Dec 2015



%\cite{Maziashvili:2012zr}
\bibitem{Maziashvili:2012zr} 
  M.~Maziashvili and L.~Megrelidze,
  %``Minimum-length deformed quantum mechanics/quantum field theory, issues, and problems,''
  PTEP {\bf 2013}, no. 12, 123B06 (2013)
  %doi:10.1093/ptep/ptt107
  [arXiv:1212.0958 [hep-th]].
  %%CITATION = doi:10.1093/ptep/ptt107;%%
  %4 citations counted in INSPIRE as of 30 Nov 2015



%\cite{Bawaj:2014cda}
\bibitem{Bawaj:2014cda} 
  M.~Bawaj {\it et al.},
  %``Probing deformed commutators with macroscopic harmonic oscillators,''
  Nature Communications 6, 7503 (2015)
 % doi:10.1038/ncomms8503
  [arXiv:1411.6410 [gr-qc]].
  %%CITATION = doi:10.1038/ncomms8503;%%
  %1 citations counted in INSPIRE as of 14 Dec 2015







%\bibitem{Kamke}
%E.~Kamke, "Differentialgleichungen - L\"osungsmethoden und L\"osungen" (Springer Fachmedien, 1983).


  
  

\bibitem{Landau}
L.~D.~ Landau and E.~M.~ Lifshitz, "Kvantovaia Mekhanika" (Moscow, Nauka, 1989).



 
%\cite{Silagadze:2009vu}
\bibitem{Silagadze:2009vu} 
  Z.~K.~Silagadze,
  %``Quantum gravity, minimum length and Keplerian orbits,''
  Phys.\ Lett.\ A {\bf 373}, 2643 (2009)
  %doi:10.1016/j.physleta.2009.05.053
  [arXiv:0901.1258 [gr-qc]].
  %%CITATION = doi:10.1016/j.physleta.2009.05.053;%%
  %14 citations counted in INSPIRE as of 30 Nov 2015












\end{thebibliography}
\end{document}